\documentclass[9pt,shortpaper,twoside,web]{ieeecolor}
\usepackage{generic}

\usepackage{amsmath,amssymb,amsfonts}
\usepackage{algorithmic}
\usepackage{graphicx}
\usepackage{textcomp}
\usepackage{subcaption}
\usepackage{placeins}
\usepackage[utf8]{inputenc}
\usepackage[switch]{lineno}
\usepackage[
  backend=biber,
  style=ieee,
  maxbibnames=1,  % 参考文献最多显示6个作者
  minbibnames=1,   % 超过后只保留1个作者 + et al.
  isbn=false
]{biblatex}
\AtEveryBibitem{%
  \clearfield{month}%
  \clearfield{day}%
}
\usepackage{caption}
\def\BibTeX{{\rm B\kern-.05em{\sc i\kern-.025em b}\kern-.08em
    T\kern-.1667em\lower.7ex\hbox{E}\kern-.125emX}}
\begin{document}
\title{Temperature Dependence of Gain and Time Resolution in LGAD Detectors}
\author{Weiyi Sun,  Mengzhao Li, Mei Zhao, and Zhijun Liang
\thanks{ This work was supported in part by the National Natural Science Foundation of China under  Grant 12275290, 12405224, 12175252, in part by the Ministry of Science and Technology of China under grant 2023YFA1605901, in part by the National Key Research and Development Program of China under Grant 2024YFA1610600}
\thanks{ Weiyi Sun, Mengzhao Li, Mei Zhao and Zhijun Liang are with the Institute of High Energy Physics, Chinese Academy of Sciences, Beijing 100049, China. (e-mail:mzli@ihep.ac.cn,zhaomei@ihep.ac.cn,liangzj@ihep.ac.cn)}
\thanks{ Weiyi Sun is also with the University of Chinese Academy of Sciences, Beijing 100049, China}
\thanks{ Mengzhao Li is also with the Spallation Neutron Source Science Center, Dongguan 523803, China}}

\maketitle

\begin{abstract}
Low-Gain Avalanche Diodes (LGADs) provide moderate internal gain and time resolutions of a few tens of picoseconds, making them a key technology for ultrafast timing in high-energy physics and beyond. However, both their gain and timing characteristics vary strongly with reverse-bias voltage and temperature. This work establishes a compact analytical framework that describes multi-temperature LGAD gain and timing behavior through an equivalent representation of the gain layer. The non-uniform multiplication region is replaced by an equivalent rectangular gain layer, from which a first-order bias-compensation relation for constant gain is derived and validated. Using multi-temperature measurements of LGADs designed by IHEP and fabricated by IME, together with an independent HPK dataset, we show that the gain-voltage curve family can be reconstructed from a reference-temperature main curve, substantially reducing characterization effort. The same idea is then extended to timing by decomposing the total time resolution into jitter and intrinsic components and representing their temperature dependences as component-wise equivalent bias offsets. The resulting framework provides a function-level description of multi-temperature LGAD time-resolution curves and offers a practical tool for calibration, operation, and reduced-density characterization of LGAD-based ultrafast timing systems.
\end{abstract}


\section{Introduction}
\label{sec:introduction}
\IEEEPARstart{A}{valanche} multiplication provides the physical basis for internal gain in semiconductor detectors. Among linear-mode avalanche devices, the low-gain avalanche diode (LGAD) has attracted considerable interest for precision timing applications. By combining a thin gain layer for controlled moderate multiplication with a thin active epitaxial layer for rapid charge collection, LGADs can achieve time resolutions of several tens of picoseconds\cite{cartigliaDesignOptimizationUltrafast2015,ferreroIntroductionUltraFastSilicon2021,moffatLowGainAvalanche2018}. Avalanche devices such as APDs, as well as SiPMs and SPADs, have also been extensively studied and deployed in photon detection, imaging, and timing applications \cite{acerbiUnderstandingSimulatingSiPMs2019b,bronziSPADFiguresMerit2016a}.

Practical deployments often require stable performance across substantially different temperatures. For instance, timing systems in high-energy physics are designed for cold operation to mitigate radiation-induced leakage current and noise; the ATLAS HGTD targets operation around 243~K \cite{casadoHighGranularityTimingDetector2022}, and the CMS Endcap Timing Layer (ETL) is similarly intended to operate below about 248~K in the HL-LHC environment \cite{ferreroCMSMTDEndcap2022}. In contrast, medical-physics and clinical beamline instrumentation is typically used close to room temperature, and LGAD-based systems have also been characterized for proton-therapy-related applications \cite{martivillarrealCharacterizationThinLGAD2023a,hellerDemonstrationLGADsCherenkov2025a}. Across these scenarios, the desired operating point generally has to be maintained by temperature compensation and multi-temperature calibration.

Extensive studies have characterized avalanche devices and LGADs as functions of temperature, bias, fabrication process, and irradiation \cite{wegrzeckaTemperatureCharacteristicsSilicon2001,ogasawaraTemperatureDependenceThin2009,zareefCharacterizationUltraFastSilicon2024,zhaoLowGainAvalanche2022b,liEffectsShallowCarbon2022,wuDesignTestingLGAD2023}. Those works establish detailed performance trends, but the resulting temperature-dependent curve families are often handled case by case. A compact relation that maps temperature variations to equivalent reverse-bias variations would therefore be valuable both physically and practically: it could clarify how gain and timing evolve with operating condition and reduce the density of repeat measurements required for calibration.

Motivated by this need, this paper develops an interpretable device-level framework for what we term bias--temperature equivalence: to first order, a temperature change can be represented as an equivalent reverse-bias offset for a fixed observable. For gain, a rectangular gain-layer equivalence (rectGL) relates bias, temperature, and gain through three device-level parameters. For timing, the same idea is extended to the jitter and intrinsic components, enabling a compact reconstruction of multi-temperature time-resolution curves. The framework is validated on measurements of LGADs designed by IHEP and fabricated by IME, together with an independent HPK dataset, and supports reduced-density characterization using a reference-temperature main curve and a small number of anchor points.

\section{Device Structure and Field Profile}
\label{sec:stru}
\begin{figure}
    \centering
    \subfloat[]{\includegraphics[width=0.75\linewidth]{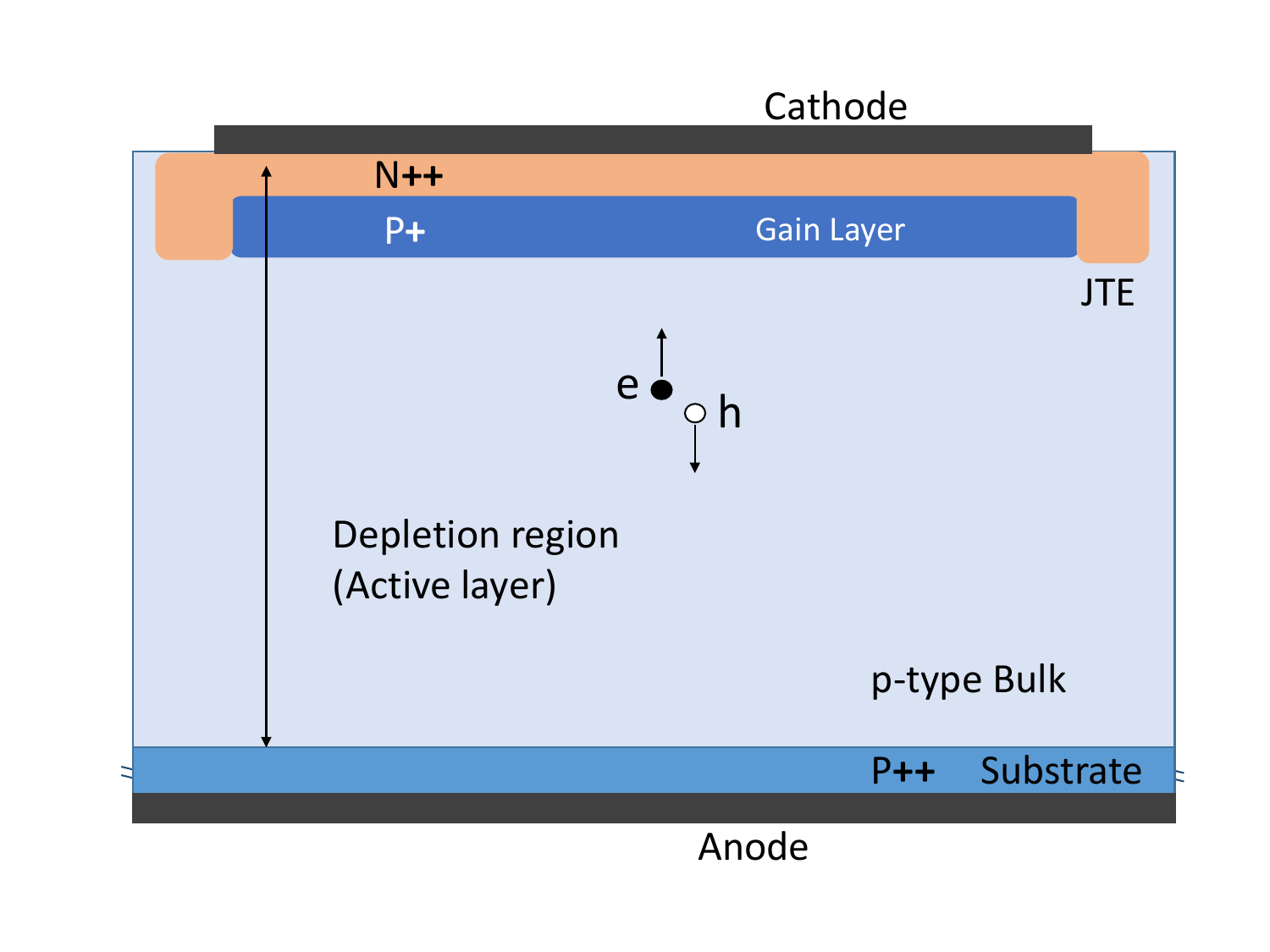}\label{fig:stru_a}}
    \hfill
    \subfloat[]{\includegraphics[width=0.75\linewidth]{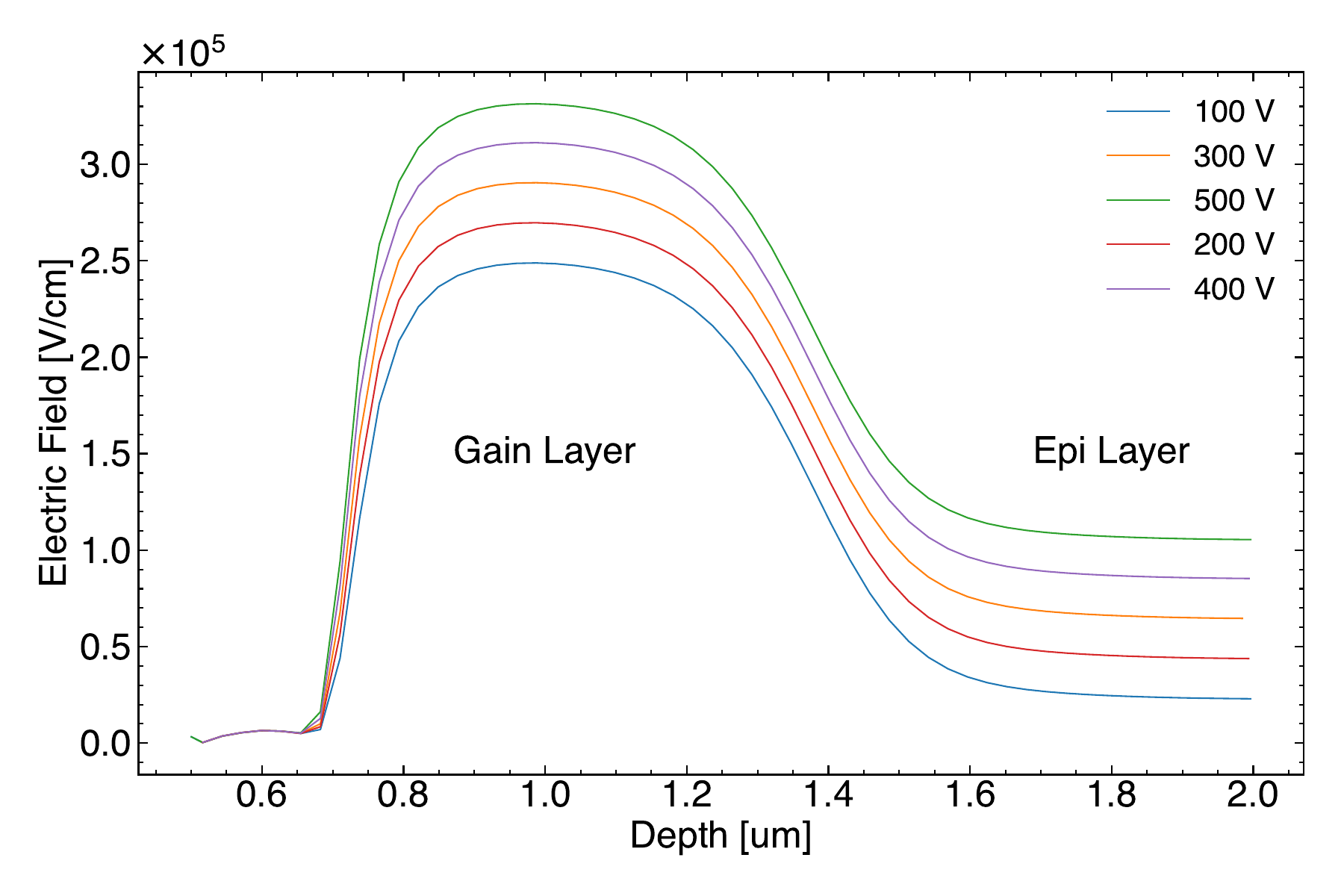}\label{fig:EF_all}}
    \hfill
    \subfloat[]{\includegraphics[width=0.75\linewidth]{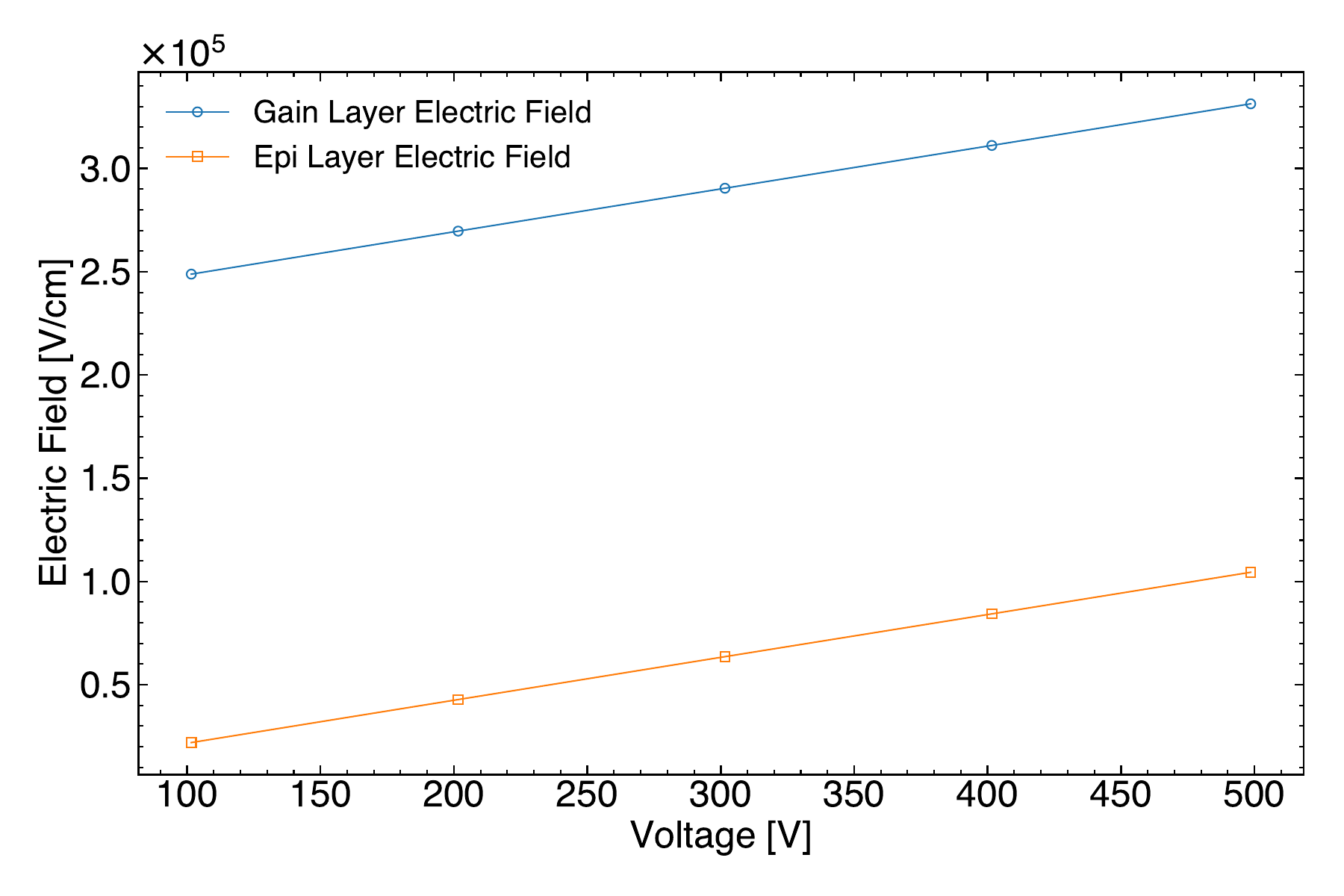}\label{fig:EF_linear}}
    \caption{(a) Schematic cross-section of the IHEP-designed, IME-fabricated LGAD sensor. (b) Simulated electric-field distribution along the depth of the LGAD structure. (c) Simulated relationship between reverse-bias voltage and characteristic electric field in the gain layer and the epitaxial layer.}
    \label{fig:stru}
\end{figure}

An LGAD is a PIN-like structure with a highly doped P$^{+}$ multiplication implant inserted near the junction, as sketched in Fig.~\ref{fig:stru}(a). The device studied here, denoted V4-R5, was designed by IHEP and fabricated by IME; for brevity, we refer to this sample family as the IHEP--IME LGAD. In this device, the gain layer is formed by 400-keV ion implantation and extends over approximately $0.5$--$1.5~\mu$m, providing a moderate gain of about 10--100. Beneath it lies a $\sim 50~\mu$m-thick $p$-type epitaxial layer grown on a heavily doped P$^{++}$ substrate with backside metallization, which serves as the back electrode and the mechanical support. Junction termination extension (JTE) structures are used to improve edge-field uniformity and suppress premature breakdown.

The simulated electric-field profile $E(x)$ is shown in Fig.~\ref{fig:stru}(b), and Fig.~\ref{fig:stru}(c) shows that, under the depletion approximation, the characteristic field in both the gain layer and the epitaxial region varies approximately linearly with reverse-bias voltage over the operating range considered here. This behavior motivates the linearized effective-field relation introduced in Section~\ref{sec:gain}.

\section{Gain Modeling Framework}
\label{sec:gain}

\subsection{Theoretical background}

When the local electric field becomes sufficiently high, carriers in silicon may gain enough energy between collisions to exceed an effective threshold energy, thereby triggering ionization and generating additional $e$-$h$ pairs. Repetition of this process in a high-field region leads to avalanche multiplication.  

Carrier acceleration in the electric field, together with phonon or impurity scattering and ionization scattering, can be described by the Boltzmann transport equation (BTE) \cite{jacoboniReviewChargeTransport1977a,conwellHighFieldTransport1981},
\begin{equation}
\frac{\partial f}{\partial t}
+\mathbf{v}(\mathbf{k})\cdot\nabla_{\mathbf{r}}f
+\frac{q\mathbf{E}}{\hbar}\cdot\nabla_{\mathbf{k}}f
=
\left(\frac{\partial f}{\partial t}\right)_{\mathrm{coll}},
\label{eq:BTE}
\end{equation}
where $f(\mathbf{r},\mathbf{k},t)$ is the phase-space distribution function and the collision term accounts for phonon or impurity scattering as well as impact ionization. In practice, the local field approximation (LFA) is commonly used: for a given local field $E$ and temperature $T$ (K), the carrier energy distribution rapidly approaches a steady state $f(\varepsilon;E,T)$. If the microscopic impact-ionization rate at energy $\varepsilon$ is $\nu_{\mathrm{ii}}(\varepsilon)$, the impact-ionization coefficient $\alpha(E,T)$, i.e., the mean number of ionization events per unit length, can be defined as
\begin{equation}
\alpha(E,T)=\frac{\langle \nu_{\mathrm{ii}}\rangle}{\langle v_d\rangle},\qquad
\langle \nu_{\mathrm{ii}}\rangle=\int \nu_{\mathrm{ii}}(\varepsilon)\,f(\varepsilon;E,T)\,d\varepsilon,
\label{eq:alpha_def}
\end{equation}
where $\langle v_d\rangle$ is the mean drift velocity. Directly solving equation~(\ref{eq:BTE}) to obtain $\alpha(E,T)$ is computationally expensive, and therefore, Chynoweth-type empirical parameterizations of $\alpha_{n,p}(E,T)$ are widely used in silicon, such as the classical forms by Massey, van Overstraeten-de Man, and Okuto-Crowell \cite{chynowethIonizationRatesElectrons1958,masseyTemperatureDependenceImpact2006,overstraetenMeasurementIonizationRates1970a,okutoIonizationCoefficientsSemiconductors1974}. 

Carrier multiplication can be formulated from the steady-state continuity equations \cite{mcintyreMultiplicationNoiseUniform1966}. Denoting the electron and hole current densities by $J_n$ and $J_p$, respectively, the impact-ionization generation rate $\mathcal{G}$ can be written as
\begin{equation}
\mathcal{G}
=\alpha_n(E,T)\frac{|J_n|}{q}+\alpha_p(E,T)\frac{|J_p|}{q},
\label{eq:Gii}
\end{equation}
together with
\begin{equation}
\frac{dJ_n}{dx}=q\mathcal{G},\qquad
\frac{dJ_p}{dx}=-q\mathcal{G}.
\label{eq:cont}
\end{equation}
Assuming a uniform-field thin region where $\alpha_n$ and $\alpha_p$ are spatial constants, and adopting the electron-injection boundary conditions
\begin{equation}
J_n(0)=J_0,\qquad J_p(d)=0,\qquad J_n(d)=J,
\label{eq:bc}
\end{equation}
where $J_0$ is the injected electron current density and $J$ is the multiplied electron current density at the exit of the high-field region, one obtains the uniform-field mean gain
\begin{equation}
G(E,T)=
\frac{\Delta\alpha\,e^{\Delta\alpha d}}
{\alpha_n-\alpha_p e^{\Delta\alpha d}},
\qquad
\Delta\alpha= \alpha_n-\alpha_p .
\label{eq:mcintyre}
\end{equation}

\subsection{Rectangular gain-layer equivalence}
In a real LGAD, the electric field is non-uniform, and can be denoted as $E(x;V)$. A rigorous gain calculation therefore requires the field profile (for example, from TCAD) together with a numerical solution of equations~(\ref{eq:Gii})--(\ref{eq:cont}) \cite{currasriveraStudyImpactIonization2023}. Guided by the field distributions in Fig.~\ref{fig:stru}(b) and Fig.~\ref{fig:stru}(c), we approximate the multiplication region by an equivalent uniform rectangle because the gain is dominated by the narrow high-field layer and the lower-field regions contribute little to the mean multiplication. Equation~(\ref{eq:mcintyre}) then becomes
\begin{equation}
G(V,T)\approx G\!\left(E_{\mathrm{eff}}(V),T;d_{\mathrm{eff}}\right).
\label{eq:G_rectGL}
\end{equation}

\begin{figure}
    \centering
    \includegraphics[width=0.75\linewidth]{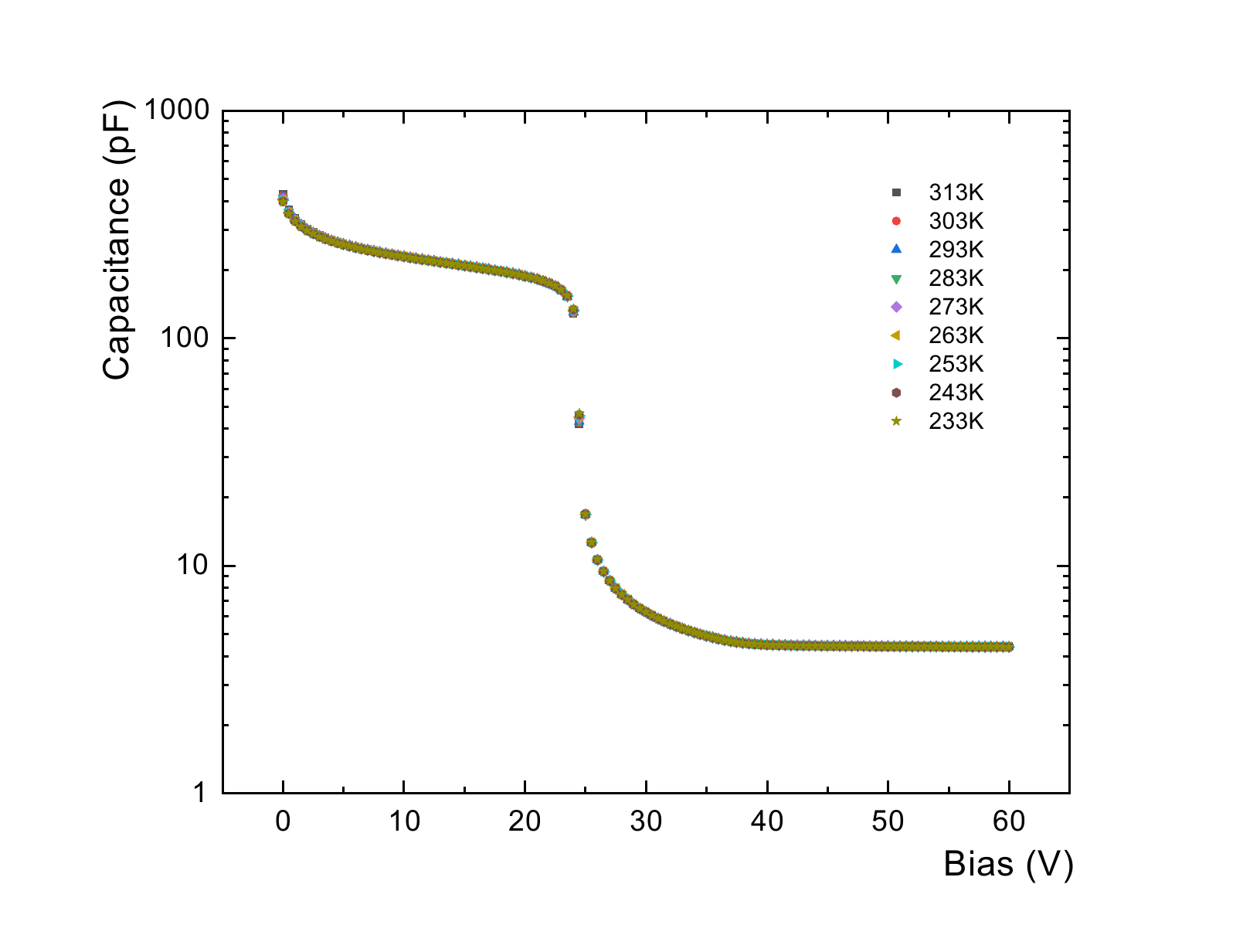}
    \caption{C-V curve of IHEP-IME LGAD measured at different temperatures.}
    \label{fig:cv}
\end{figure}

The resulting rectangular gain-layer equivalence (rectGL) is a one-dimensional mean-gain model that reproduces the measured average gain. Using the approximately linear field--voltage relation discussed in Section~\ref{sec:stru}, we linearize the effective field in the multiplication region as
\begin{equation}
E_{\mathrm{eff}}(V)=E_0+s\,(V-V_{\mathrm{gl}}),\qquad
s= \frac{dE_{\mathrm{eff}}}{dV},
\label{eq:Eeff_linear}
\end{equation}
where $E_0$ is a reference field, $V_{\mathrm{gl}}$ is the reverse-bias voltage at which the gain layer becomes fully depleted, and $s$ is an effective field-to-voltage conversion coefficient determined by the depleted thickness and doping profile. For the IHEP--IME prototypes studied here, the measured C--V curves (shown in fugure \ref{fig:cv}) are nearly identical from 233 to 313~K, and the extracted $V_{\mathrm{gl}}$ and full-depletion voltage $V_\mathrm{fd}$ of the full sensor are $24.6 \pm 0.3$~V and $45.7 \pm 0.1$~V, respectively. This supports the assumption that the depletion boundaries are effectively pinned in this temperature range. We therefore treat $E_{\mathrm{eff}}(V)$ as temperature-independent to first order and attribute the dominant temperature dependence to $\alpha_{n,p}(E,T)$. Once a specific $\alpha_{n,p}(E,T)$ parameterization is chosen, the $G$--$V$ relation is determined by the three device-level equivalent parameters $(E_0,s,d_{\mathrm{eff}})$.

For a prescribed target gain $G_0$, we define the equal-gain residual function
\begin{equation}
F(V,T;G_0)= \ln G(V,T)-\ln G_0=0.
\label{eq:F_def}
\end{equation}
Implicit differentiation gives the temperature-compensation slope
\begin{equation}
k(G_0)= \frac{dV}{dT}\Big|_{G_0}
=-\frac{\partial_T\ln G}{\partial_V\ln G}
=-\frac{\partial_T\ln G}{s\,\partial_E\ln G}.
\label{eq:k_def}
\end{equation}
Thus, around a reference temperature $T_0$, one obtains the first-order approximation
\begin{equation}
V(G_0,T)\approx V(G_0,T_0)+k(G_0)(T-T_0).
\label{eq:VT_linear}
\end{equation}

To illustrate why $k(G_0)$ depends only weakly on $G_0$, consider the electron-dominated limit ($\alpha_p\ll\alpha_n$) in which $\ln G\approx \alpha_n d_{\mathrm{eff}}$. If the standard Massey parameterization with the literature coefficients reported in \cite{masseyTemperatureDependenceImpact2006} is used,
\begin{equation}
\alpha_n(E,T)=A_n\exp\!\left[-\frac{C_n+D_n T}{E}\right]\qquad (T\ \mathrm{in\ K}).
\label{eq:massey}
\end{equation}
For fixed $G_0$, $\alpha_n=\ln G_0/d_{\mathrm{eff}}$ leads to
\begin{equation}
k(G_0)=\left.\frac{dV}{dT}\right|_{G_0}=\frac{D_n}{s\,\ln\!\left(\frac{A_n d_{\mathrm{eff}}}{\ln G_0}\right)} .
\label{eq:k_massey}
\end{equation}
Equation~(\ref{eq:k_massey}) predicts an exactly linear $V$--$T$ relation within this approximation and a logarithmically weak dependence of $k$ on $G_0$. In practice, all fits and predictions in this work are obtained numerically from equation~(\ref{eq:mcintyre}) with the chosen $\alpha_{n,p}(E,T)$ parameterization.

\subsection{Validation of the rectGL framework}
\subsubsection{Validation on IHEP--IME multi-temperature data}
\begin{figure}[htbp]
    \centering
    \subfloat[]{\includegraphics[width=0.8\linewidth]{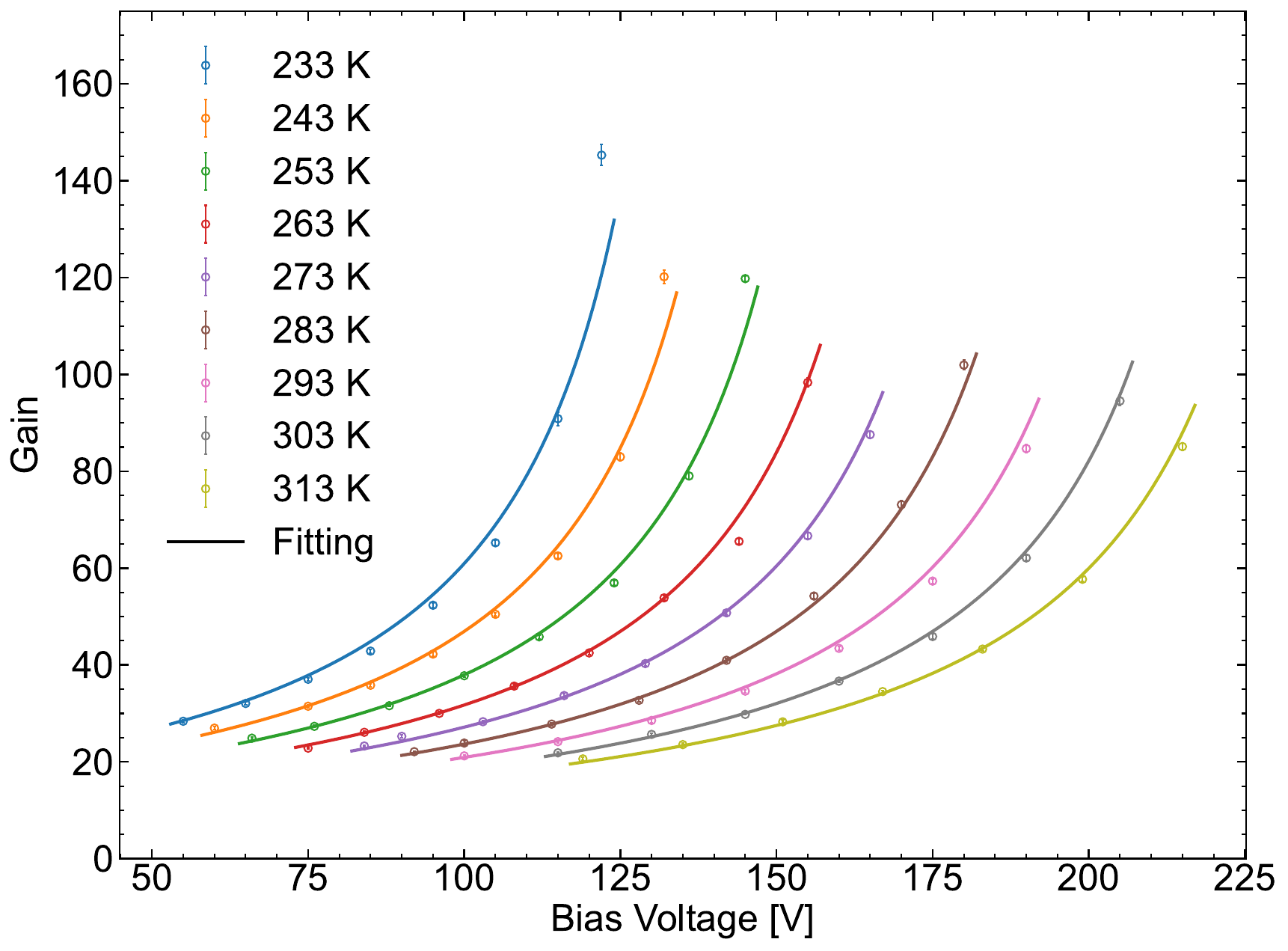}\label{fig:9_full}}
    \hfill
    \subfloat[]{\includegraphics[width=0.8\linewidth]{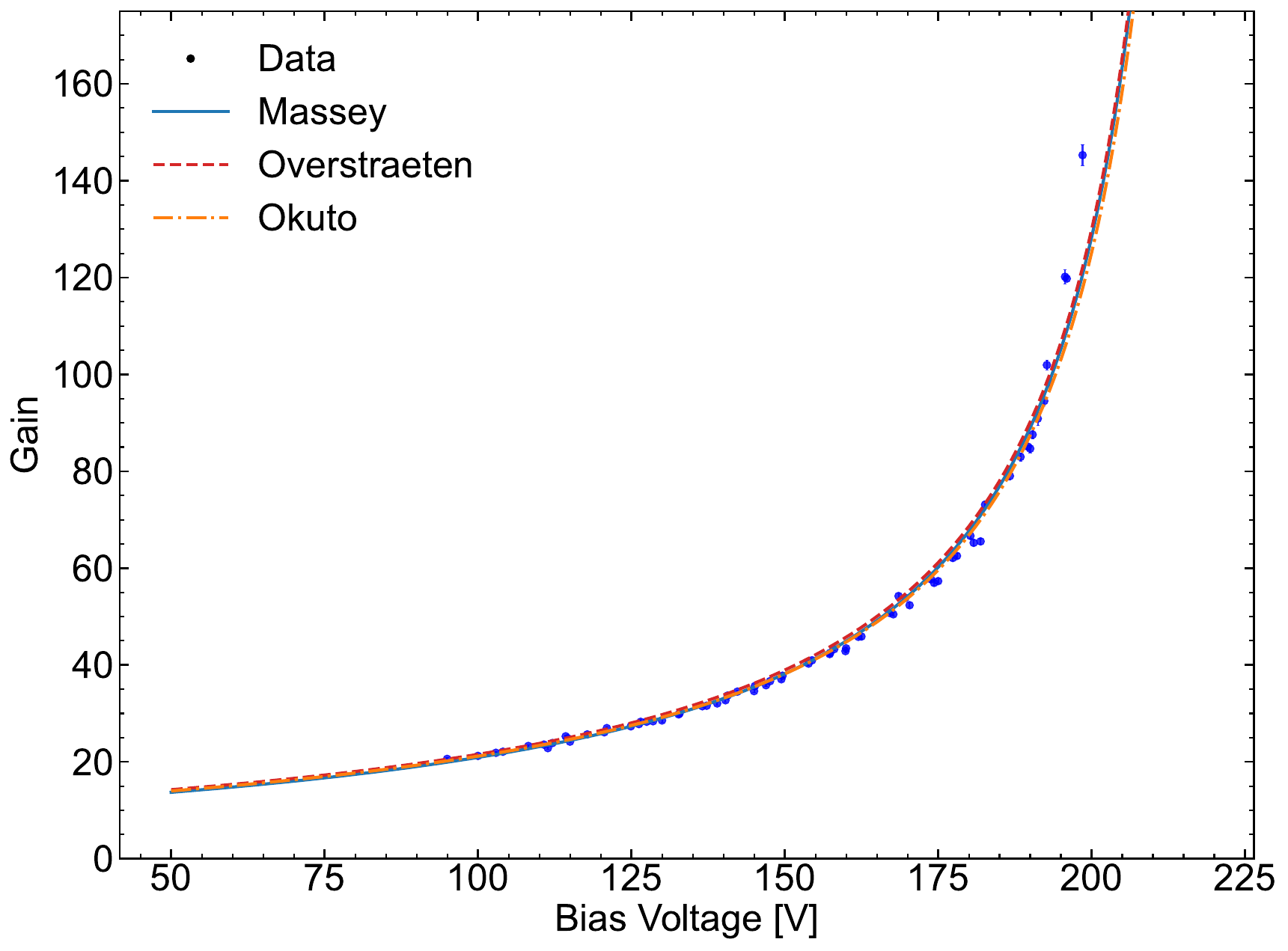}\label{fig:9_collapse}}
    \hfill
    \subfloat[]{\includegraphics[width=0.8\linewidth]{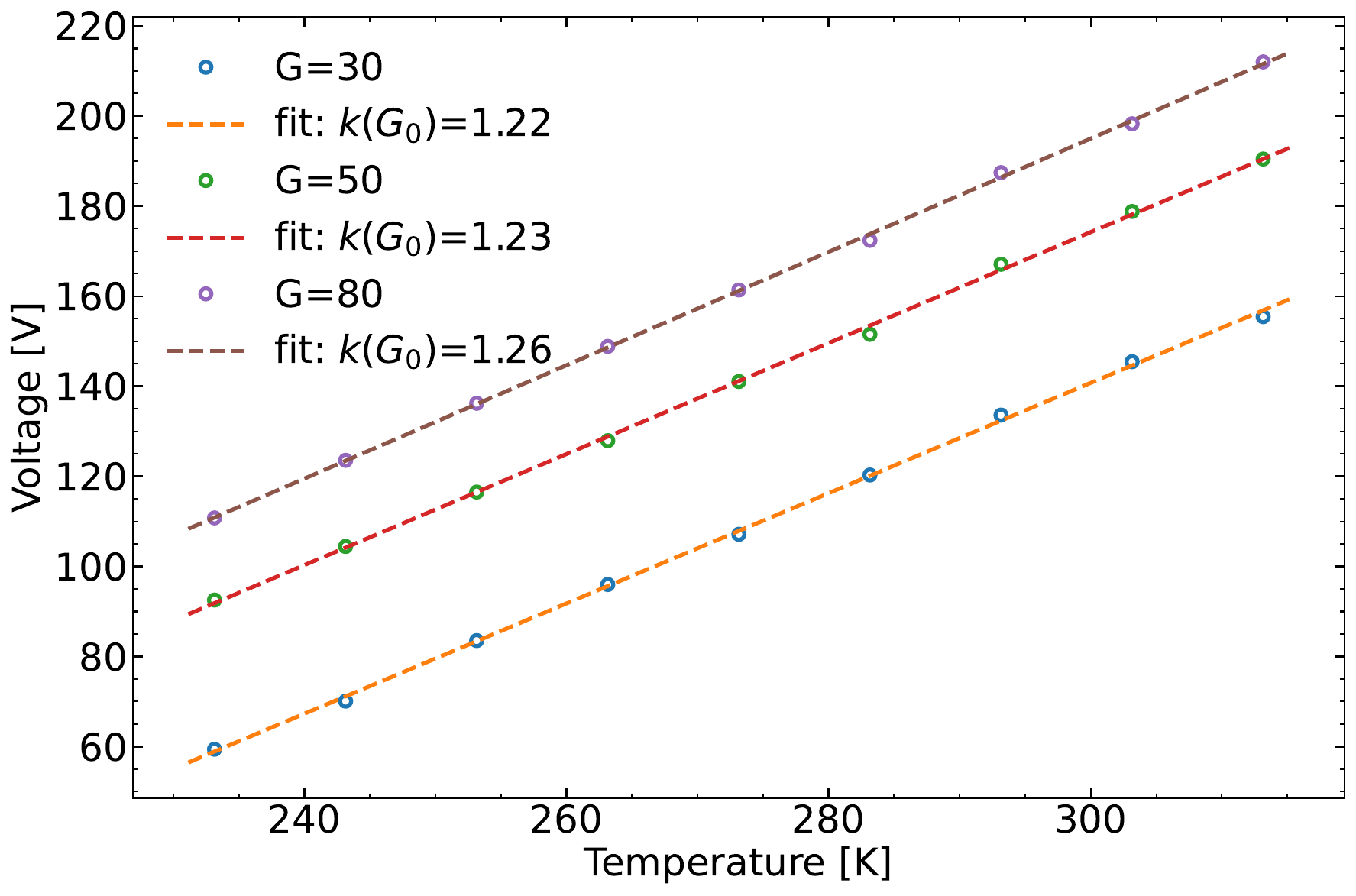}\label{fig:9_isogain}}
    \caption{Gain-side validation of the rectGL framework using the IHEP--IME dataset. (a) Global fit of the measured $G$--$V$ curves at nine temperatures. (b) Curves translated to the reference temperature of 273~K and fitted with three $\alpha(E,T)$ parameterizations. (c) Equal-gain bias values as functions of temperature for several target gains.}
    \label{fig:ihep_gain_validation}
\end{figure}

We measured the $G$--$V$ curve family of the IHEP--IME LGAD at nine temperatures between 233 and 313~K. The gain $G$ is defined as the ratio of the collected charge in the LGAD to that measured in a no-gain reference PIN device. Using the three-parameter rectGL model constructed from equations~(\ref{eq:mcintyre}) and~(\ref{eq:Eeff_linear}), we performed a global fit to all nine temperature curves, as shown in Fig.~\ref{fig:ihep_gain_validation}(a), and obtained good agreement across the full operating range.

Choosing $T_{\mathrm{ref}}=273$~K, we translate all measured curves to the reference temperature by using the bias-compensation relation implied by equation~(\ref{eq:VT_linear}). The translated data are then fitted under three different $\alpha(E,T)$ parameterizations, as shown in Fig.~\ref{fig:ihep_gain_validation}(b). Using the Massey form as an example, the fit yields
\begin{equation}
E_0= 3.1\times10^5~\mathrm{V/cm},\quad
s= 195~\mathrm{(V/cm)/V},\quad
d_{\mathrm{eff}}= 1.28~\mu\mathrm{m}.
\end{equation}
 From the nine measured curves we also extract several equal-gain lines $V(G_0,T)$ and fit them linearly. Equation~(\ref{eq:VT_linear}) is well satisfied across the full temperature range, and the slope $k(G_0)$ varies only weakly with $G_0$, as shown in Fig.~\ref{fig:ihep_gain_validation}(c). Equation~(\ref{eq:k_massey}) further indicates that a larger effective ionization path length $d_{\mathrm{eff}}$ weakens the gain dependence of $k(G_0)$, providing a compact process-level interpretation of the measured trend.

\begin{figure}[htbp]
    \centering
    \includegraphics[width=0.8\linewidth]{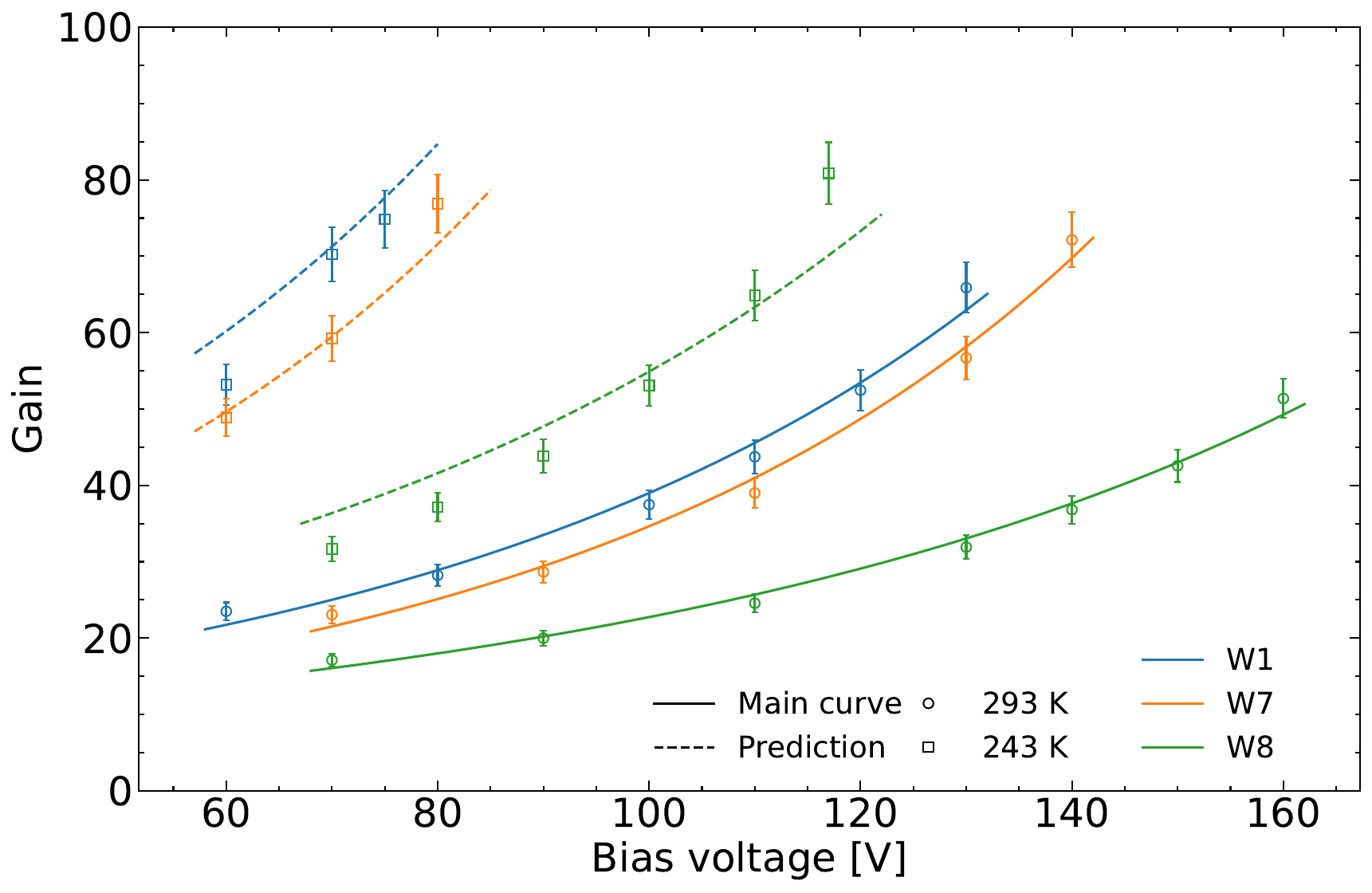}
    \caption{Application of the rectGL framework to three previously reported IHEP--IME LGAD samples measured at 243~K and 293~K. Solid lines denote fits to the reference curves, and dashed lines denote predictions at the second temperature. Data are taken from \cite{zhaoLowGainAvalanche2022b}.}
    \label{fig:ihep_legacy_collapse}
\end{figure}

We further test the same framework on three previously published IHEP--IME LGAD samples from wafers W1, W7, and W8 reported in \cite{zhaoLowGainAvalanche2022b}, as shown in Fig.~\ref{fig:ihep_legacy_collapse}. These wafers correspond to different gain-layer process conditions. For W7, the phosphorus and boron doping profiles were designed to match those of the V4-R5 sample, so $d_{\mathrm{eff}}$ was fixed and the 293~K reference curve was fitted using two points; the 243~K curve was then predicted without additional fit parameters. For W1 and W8, all three rectGL parameters were fitted to the reference curve and used to predict the second-temperature curve. The extracted $d_{\mathrm{eff}}$ values for W1, W7, and W8 are 1.71, 1.28, and 1.16~$\mu$m, respectively. These values follow the known process variations: W1 includes carbon implantation in the gain layer, whereas W8 uses a higher $N^{++}$ implantation energy and a narrower gain layer. This supports the interpretation that the fitted effective gain layer captures meaningful process-dependent information about the real multiplication region.

\subsubsection{Transferability to the HPK dataset}
\begin{figure}[htbp]
    \centering
    \subfloat[]{\includegraphics[width=0.85\linewidth]{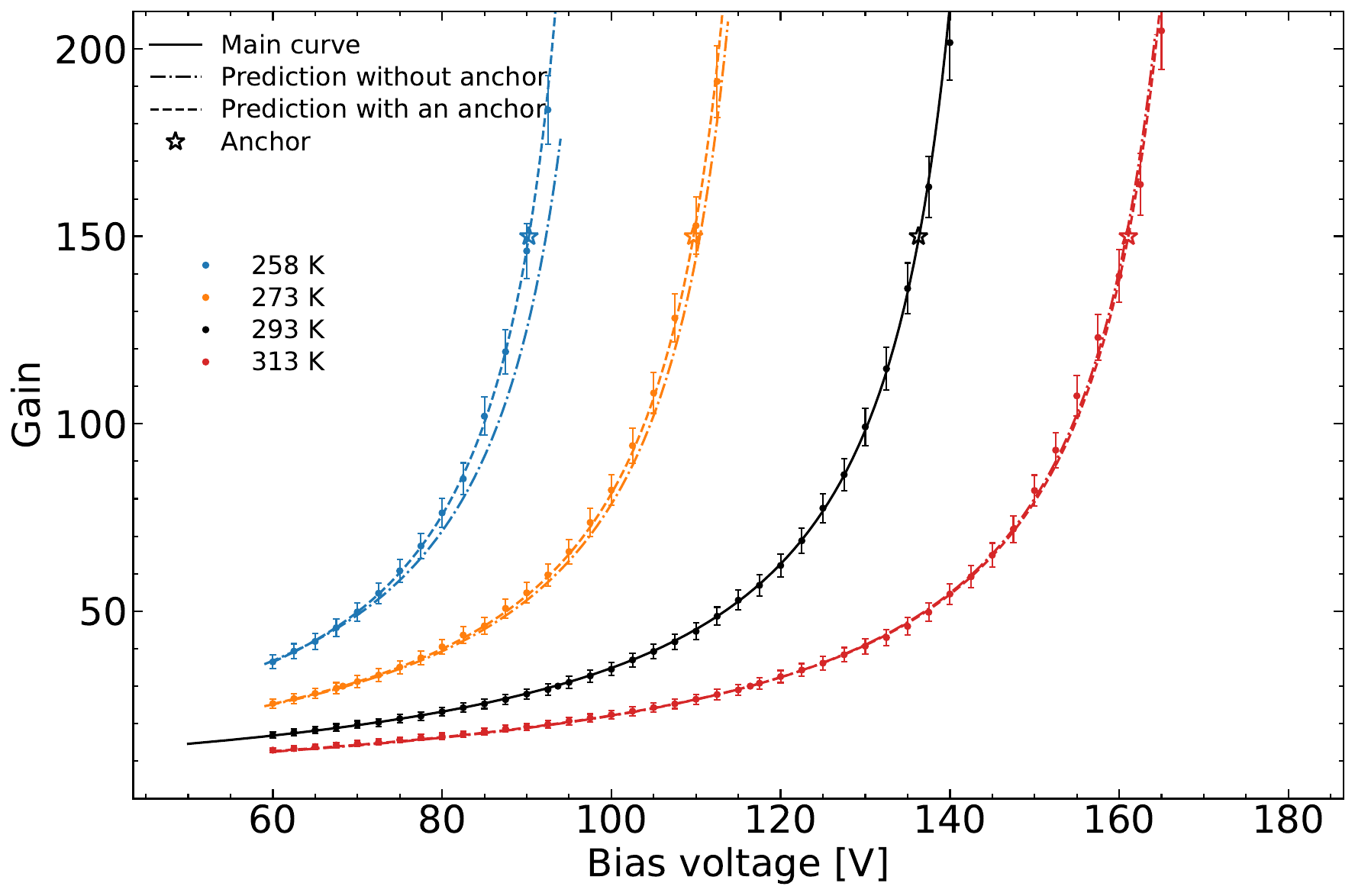}\label{fig:cern_collapse}}
    \hfill
    \subfloat[]{\includegraphics[width=0.85\linewidth]{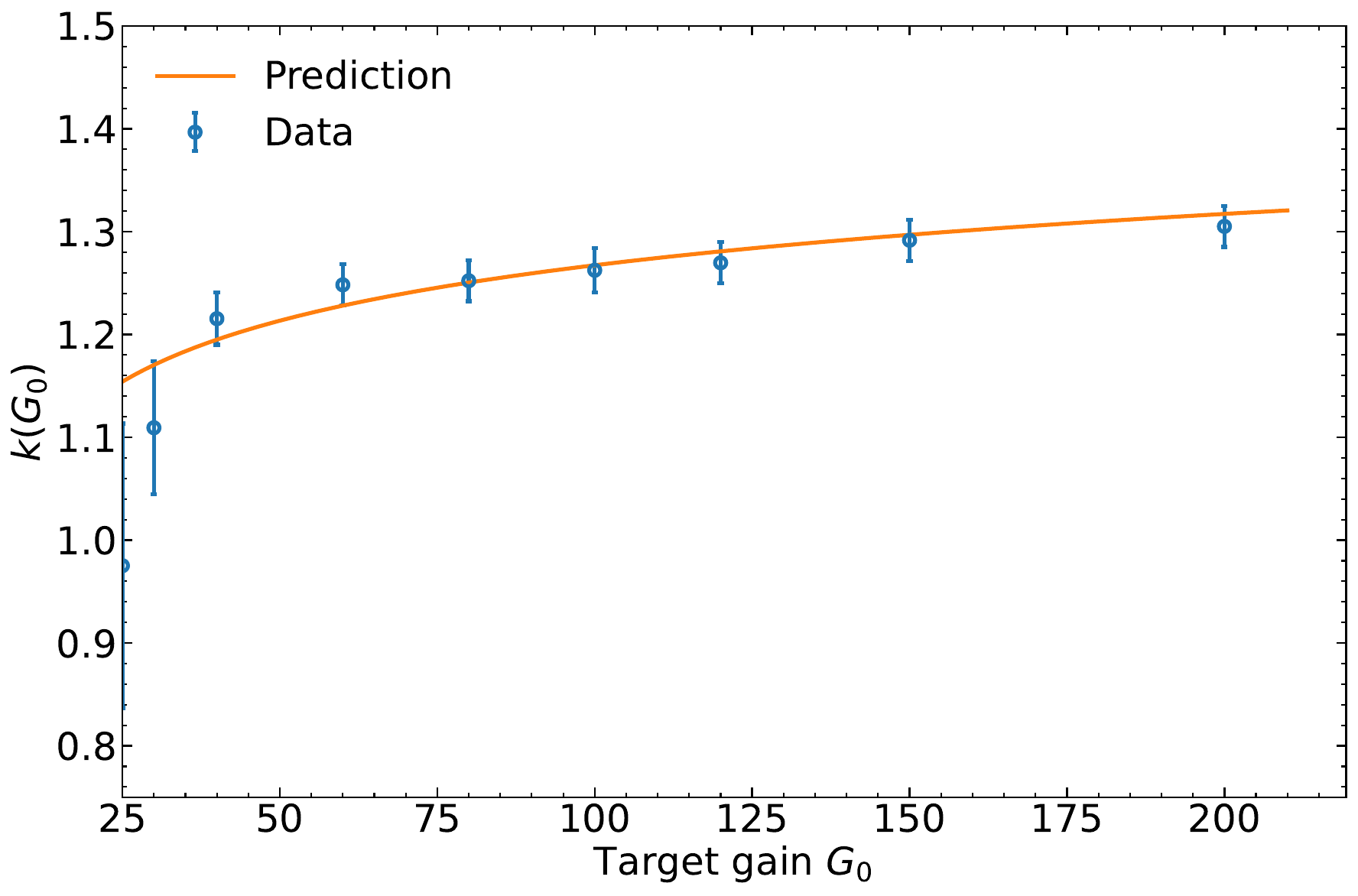}\label{fig:cern_kg0}}
    \hfill
    \subfloat[]{\includegraphics[width=0.85\linewidth]{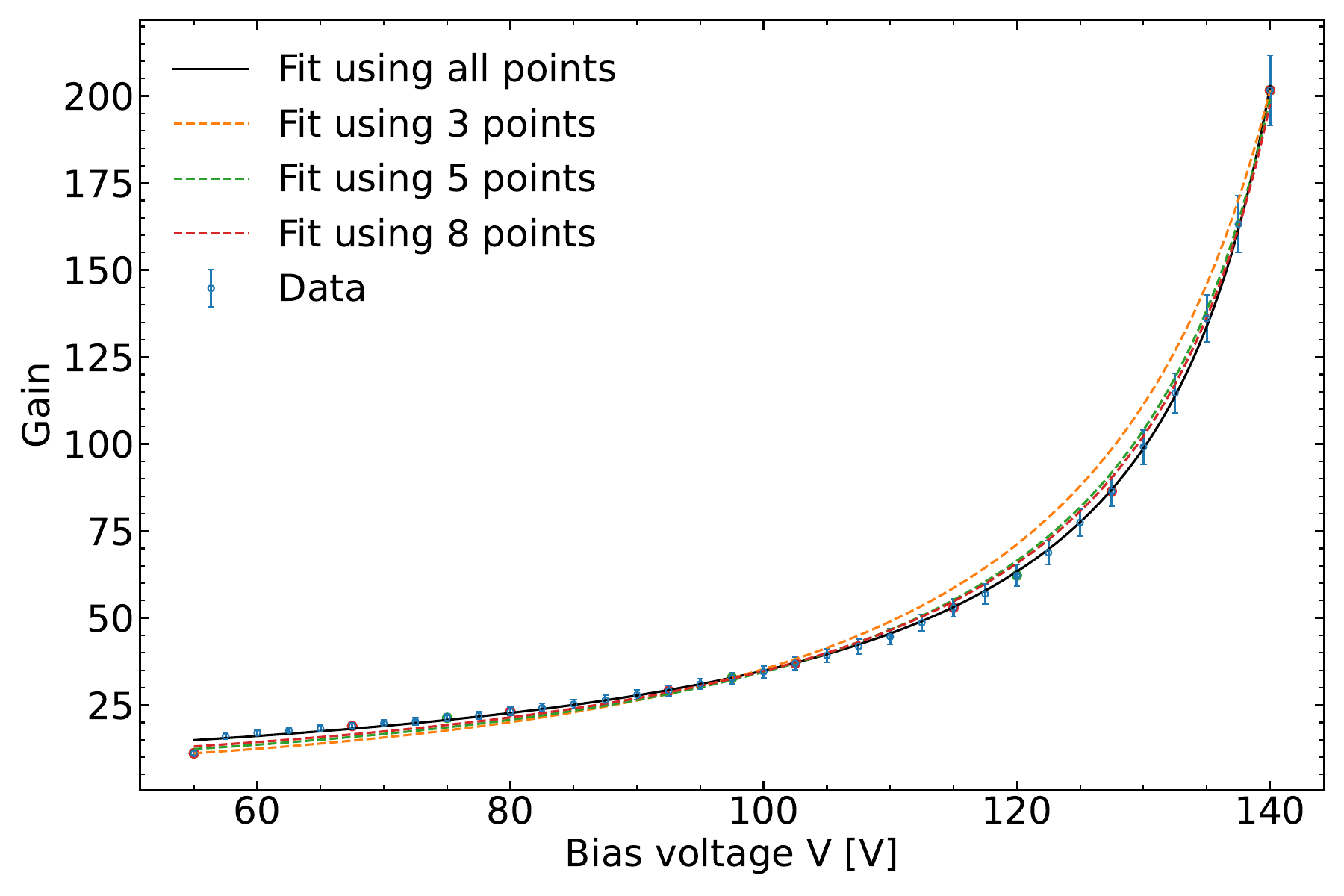}\label{fig:9_pts}}
    \caption{Gain-side transferability test using the HPK dataset from \cite{currasriveraStudyImpactIonization2023}. (a) Fit to the 293~K reference curve and predictions at other temperatures, with anchor-corrected predictions also shown. (b) Temperature-compensation slope $k(G_0)$ as a function of target gain. (c) Reconstruction of the reference main curve using different numbers of sampled bias points.}
    \label{fig:hpk_validation}
\end{figure}

To test transferability, we apply the same procedure to the published multi-temperature data of the HPK2-S1 LGAD \cite{currasriveraStudyImpactIonization2023}. The rectGL model is first fitted to the reference main curve $G_{\mathrm{main}}(V;T_0)$ at $T_0=293$~K. With those parameters fixed, the $G$--$V$ curves at the other temperatures are predicted, as shown in Fig.~\ref{fig:hpk_validation}(a). The overall agreement is good, although the deviations become larger as the temperature moves farther from $T_0$. The extracted $k(G_0)$ again shows only a weak dependence on target gain, as summarized in Fig.~\ref{fig:hpk_validation}(b).

The residual discrepancy suggests that the effective $E$--$V$ relation may retain a weak temperature dependence beyond the first-order approximation. To account for this effect, we use one measured point at each temperature as an anchor to update the effective slope $s$ in equation~(\ref{eq:Eeff_linear}). The anchor-corrected curves, also shown in Fig.~\ref{fig:hpk_validation}(a), agree better with the measurements.

This leads to a practical characterization workflow: (i) fit the reference-temperature main curve; (ii) predict the full multi-temperature family from the fitted rectGL parameters; and (iii) if higher accuracy is needed, use one anchor point per temperature to correct the residual drift of the effective field-to-voltage coefficient.

To reduce repeat-testing cost, we also evaluate how many bias points are needed to reconstruct the reference main curve, as shown in Fig.~\ref{fig:hpk_validation}(c). Although the three-parameter model is mathematically determined by three points, the reconstruction becomes sensitive to point placement and noise in that limit. With five points spanning the low-, mid-, and high-gain regions, the main-curve error can be kept near the 10\% level; with eight points, it decreases toward 7\%. We therefore recommend measuring 5--8 bias points at the reference temperature and then reconstructing the multi-temperature family with one or two anchor points per additional temperature.

\section{Bias--Temperature Equivalence for Time Resolution}
\label{sec:timing_shift}

The time resolution of an LGAD is taken as the standard deviation $\sigma_t$ of the measured arrival-time distribution under fixed experimental conditions. In this section, $V_{\rm b}$ denotes the applied reverse-bias voltage.

For the measurements considered here, the dominant contributions to the total time resolution can be written as \cite{cartigliaDesignOptimizationUltrafast2015,ferreroIntroductionUltraFastSilicon2021}
\begin{equation}
\sigma_t^2(T,V_{\rm b})
=
\sigma_{\rm Landau}^2
+\sigma_{\rm jitter}^2
+\sigma_{\rm TW}^2
+\sigma_{\rm dist}^2,
\label{eq:sigt_decomp}
\end{equation}
where $\sigma_{\rm Landau}$ denotes the charge-deposition fluctuation term, $\sigma_{\rm TW}$ the residual time-walk contribution, and $\sigma_{\rm dist}$ a compact residual term that collects waveform distortion and other weak systematic effects. After constant-fraction-discriminator (CFD) correction, the time-walk term is strongly suppressed. For the present dataset, the remaining systematic terms are small compared with the dominant jitter and intrinsic contributions, so they are absorbed into the residual intrinsic term defined below. We therefore use the working approximation
\begin{equation}
\sigma_t^2(T,V_{\rm b})
\approx
\sigma_{\rm jitter}^2(T,V_{\rm b})
+\sigma_{\rm int}^2(T,V_{\rm b}),
\label{eq:sig_quadsum}
\end{equation}
where $\sigma_{\rm jitter}$ denotes the timing jitter and $\sigma_{\rm int}$ is the residual intrinsic component after subtracting the jitter term.

% ----------------------------------------------------------------------
\subsection{Jitter component}
\label{subsec:jitter_shift}

A commonly used expression for jitter is \cite{ferreroIntroductionUltraFastSilicon2021}
\begin{equation}
\sigma_{\rm jitter}(T,V_{\rm b})
\approx
k_j\,\frac{t_r(T,V_{\rm b})}{\mathrm{SNR}(T,V_{\rm b})},
\label{eq:jitter_def}
\end{equation}
where $t_r$ is the signal rise time, $\mathrm{SNR}$ is the signal-to-noise ratio, and $k_j$ is a constant factor determined by the discriminator choice and waveform shape.

Assuming that the rising edge is dominated by carrier drift in the epitaxial layer, one has $t_r=d_{\rm epi}/v_d$, where $d_{\rm epi}$ is the epitaxial thickness and $v_d$ is the longitudinal drift velocity. We model $v_d$ with the empirical Caughey--Thomas (CT) relation \cite{caugheyCarrierMobilitiesSilicon1967,jacoboniReviewChargeTransport1977a}:
\begin{equation}
v_d(E,T)=\frac{\mu_0(T)\,E}{\left[1+\left(\frac{E}{E_c(T)}\right)^{\beta(T)}\right]^{1/\beta(T)}}.
\label{eq:vd_CT}
\end{equation}
Here $\mu_0(T)$, $E_c(T)$, and $\beta(T)$ are temperature-dependent empirical parameters. Using the same effective-field relation as in equation~(\ref{eq:Eeff_linear}), we define
\begin{equation}
\begin{aligned}
y(E,T)&=\left(\frac{E}{E_c(T)}\right)^{\beta(T)},\\
t_r(T,V_{\rm b})&=\frac{d_{\rm epi}}{\mu_0(T)\,E}\,(1+y)^{1/\beta(T)}.
\end{aligned}
\label{eq:tr_CT_closed}
\end{equation}
At fixed $T$, both $E_c$ and $\beta$ are constants and $d\ln y/d\ln E=\beta(T)$, which gives
\begin{equation}
\left(\frac{\partial \ln t_r}{\partial \ln E}\right)_T=-\frac{1}{1+y}.
\label{eq:dlntr_dlnE}
\end{equation}
Equation~(\ref{eq:dlntr_dlnE}) shows that the rise-time sensitivity to field weakens at higher field, allowing its bias and temperature dependence to be reconstructed compactly.

On the other hand, the signal-to-noise ratio can be written as
\begin{equation}
\mathrm{SNR}(T,V_{\rm b})
\propto
\frac{Q(T,V_{\rm b})}{\sigma_N(T,V_{\rm b})}
=
\frac{Q_0\,G(T,V_{\rm b})}{\sigma_N(T,V_{\rm b})},
\end{equation}
where $Q_0$ is the primary deposited charge and $\sigma_N$ is the equivalent noise charge. For the LGAD sample and measurement setup considered here, the measured $\sigma_N$ varies by less than 2\% over the explored bias and temperature ranges, so it is treated as constant to first order. Over the fitted bias interval, $\ln G(T,V_{\rm b})$ is well approximated by a linear function of $V_{\rm b}$, which leads to the effective representation
\begin{equation}
\ln \mathrm{SNR}(T,V_{\rm b})
\approx
A_{\rm SNR}(T)+\kappa\,V_{\rm b},
\label{eq:lnsnr_linear}
\end{equation}
where $A_{\rm SNR}(T)$ is a temperature-dependent intercept that collects $Q_0$, the nearly constant noise term, and the offset part of $\ln G$, while $\kappa$ is the effective slope in the considered bias range.

\begin{figure}[t]
\centering
\subfloat[]{\includegraphics[width=0.9\linewidth]{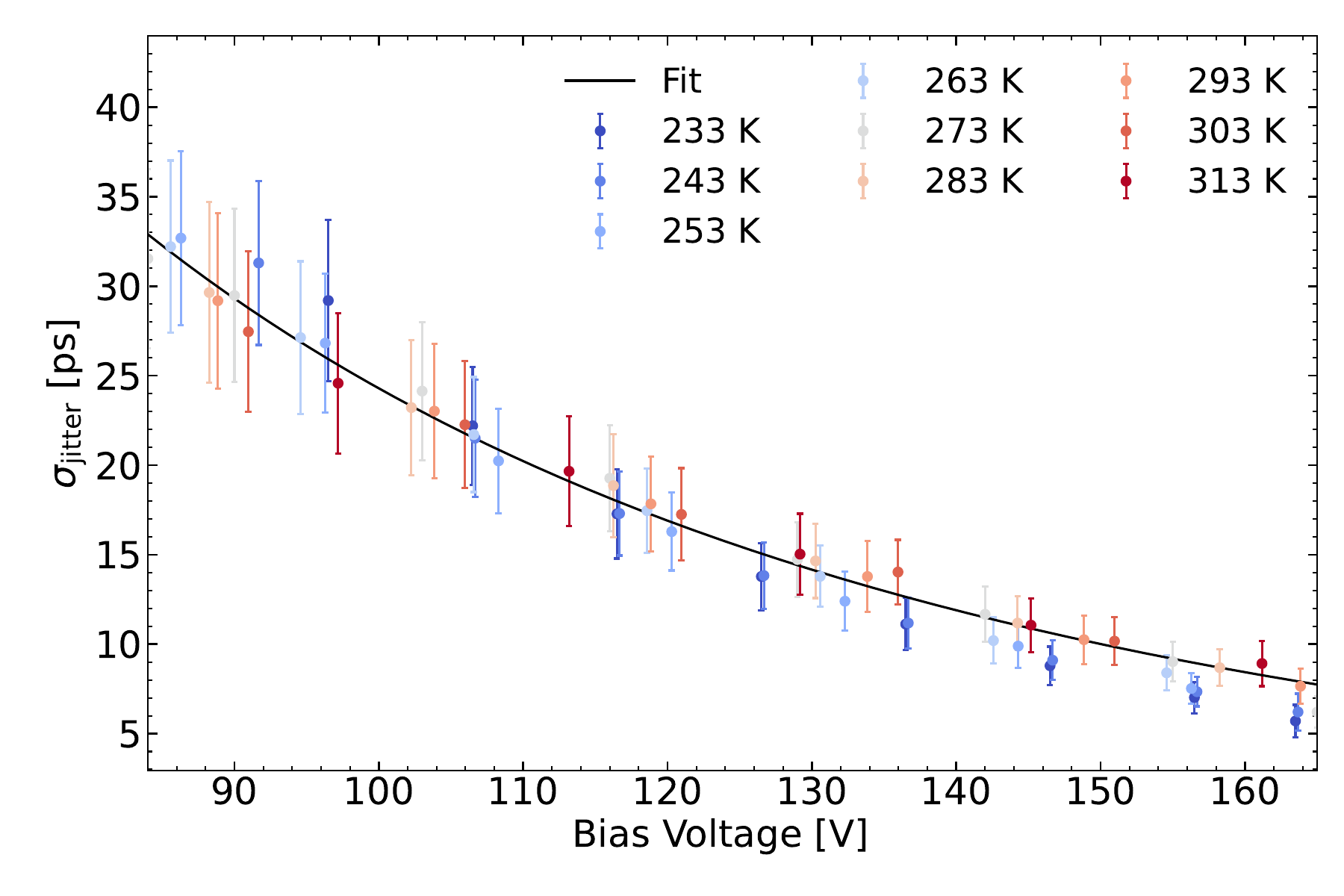}\label{fig:9_jc}}
\hfill
\subfloat[]{\includegraphics[width=0.9\linewidth]{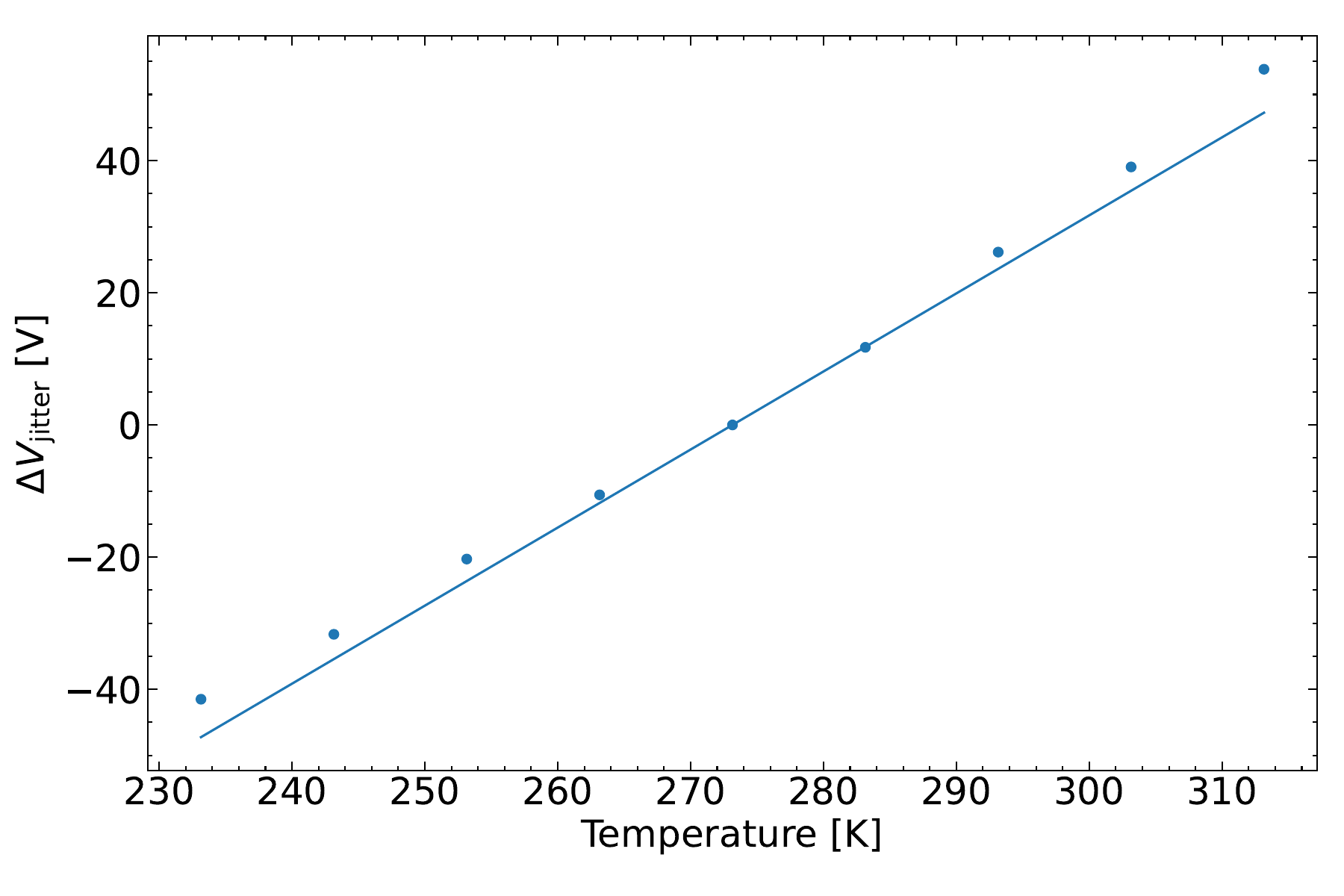}\label{fig:9_js}}
\hfill
\caption{Jitter-side bias--temperature equivalence. (a) Measured $\sigma_{\rm jitter}(V_{\rm b})$ curves at nine temperatures, translated to 273~K for the global fit. (b) Extracted equivalent bias offset $\Delta V_{\rm jitter}(T)$ with a linear fit.}
\label{fig:jitter_shift}
\end{figure}

Combining equations~(\ref{eq:jitter_def}),~(\ref{eq:dlntr_dlnE}), and~(\ref{eq:lnsnr_linear}), we represent the temperature dependence of $\sigma_{\rm jitter}$ by an equivalent bias offset. Choosing $T_{\rm ref}=273$~K and using power-law forms for $\mu_0(T)$ and $E_c(T)$ with a linear form for $\beta(T)$, we perform a global fit of $\sigma_{\rm jitter}(T,V_{\rm b})$ using the nine temperature datasets and then translate the curves along the bias axis to extract $\Delta V_{\rm jitter}(T)$, as shown in Fig.~\ref{fig:jitter_shift}(a). The compensation slope at $\sigma_{\rm jitter}=20$~ps is
\begin{equation}
\Delta V_{\rm jitter}(T)= s_{\rm jitter}(T-T_{\rm ref}),
\end{equation}
with
\begin{equation}
    s_{\rm jitter}=(1.17\pm0.03)\ {\rm V/K}.
\label{eq:shift_jitter_fit}
\end{equation}
As shown in Fig.~\ref{fig:jitter_shift}(b), $\Delta V_{\rm jitter}(T)$ exhibits a clear linear dependence over 233--313~K, with no significant nonlinearity observed.

% ----------------------------------------------------------------------
\subsection{Intrinsic component}
\label{subsec:int_shift}

The intrinsic term is defined as the residual after subtracting jitter from the total time resolution. Within the centroid timing-fluctuation framework, error-propagation and statistical-fluctuation approximations lead to $\mathrm{Var}(\hat t)\propto 1/Q$, with $Q$ the collected charge \cite{rieglerTimeResolutionSilicon2017}. We therefore use the parameterization
\begin{equation}
\sigma_{\rm int}^2(T,V_{\rm b})
=
\sigma_0^2+\frac{A_Q}{Q(T,V_{\rm b})}.
\label{eq:sigint_Q}
\end{equation}
Here $\sigma_0$ denotes the irreducible floor associated with geometry, drift-time distribution, waveform shape, and other weakly bias-dependent effects, while $A_Q$ parameterizes the charge-dependent intrinsic fluctuations.

For a fully depleted epitaxial layer and approximately temperature-independent primary charge $Q_0$, one can write $Q(T,V_{\rm b})= Q_0\,G(T,V_{\rm b})$, which yields
\begin{equation}
\sigma_{\rm int}^2(T,V_{\rm b})
=
\sigma_0^2+\frac{A_G}{G(T,V_{\rm b})},
\qquad
A_G= A_Q/Q_0 .
\label{eq:sigint_G}
\end{equation}

\begin{figure}[t]
\centering
\subfloat[]{\includegraphics[width=0.9\linewidth]{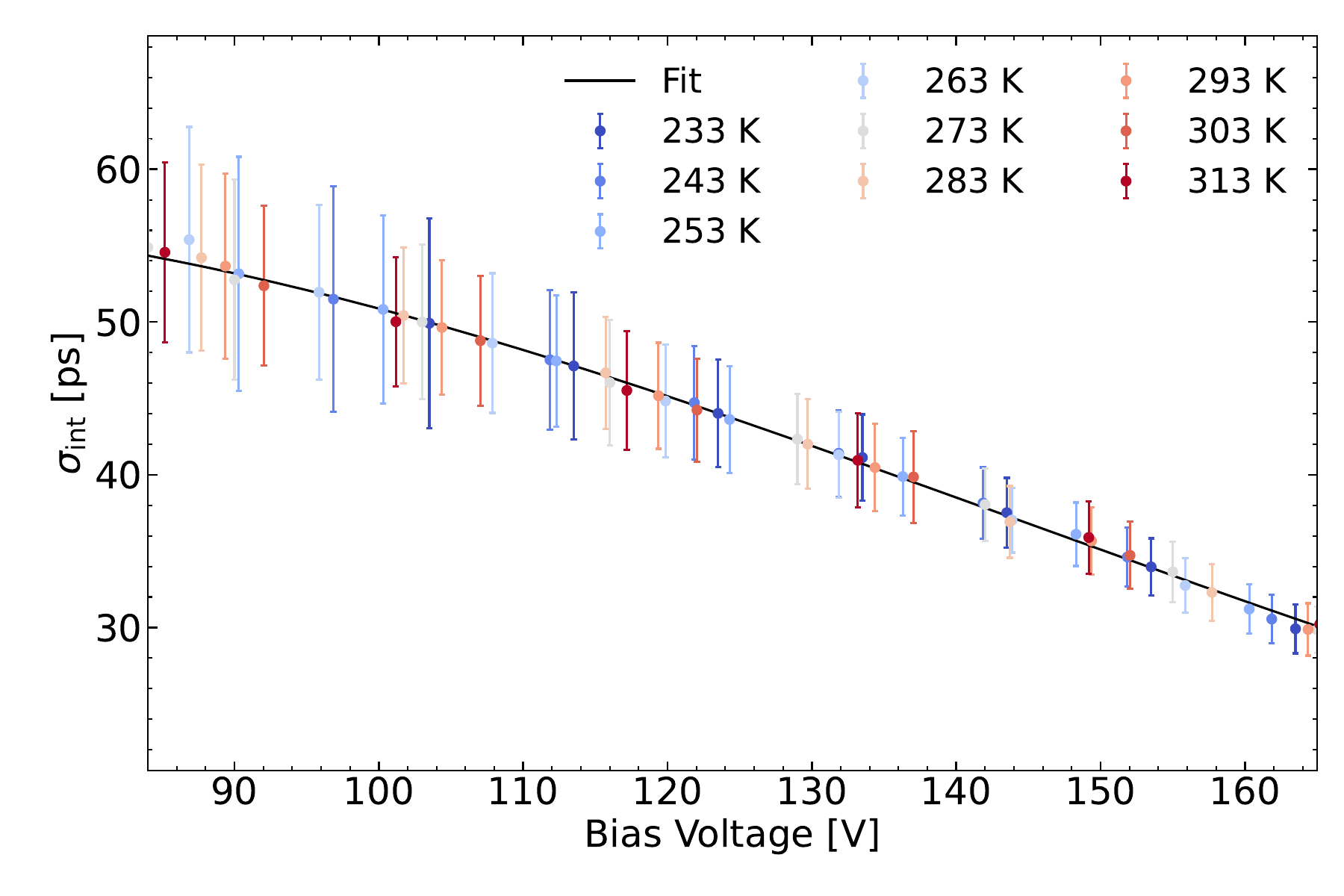}\label{fig:9_ic}}
\hfill
\subfloat[]{\includegraphics[width=0.9\linewidth]{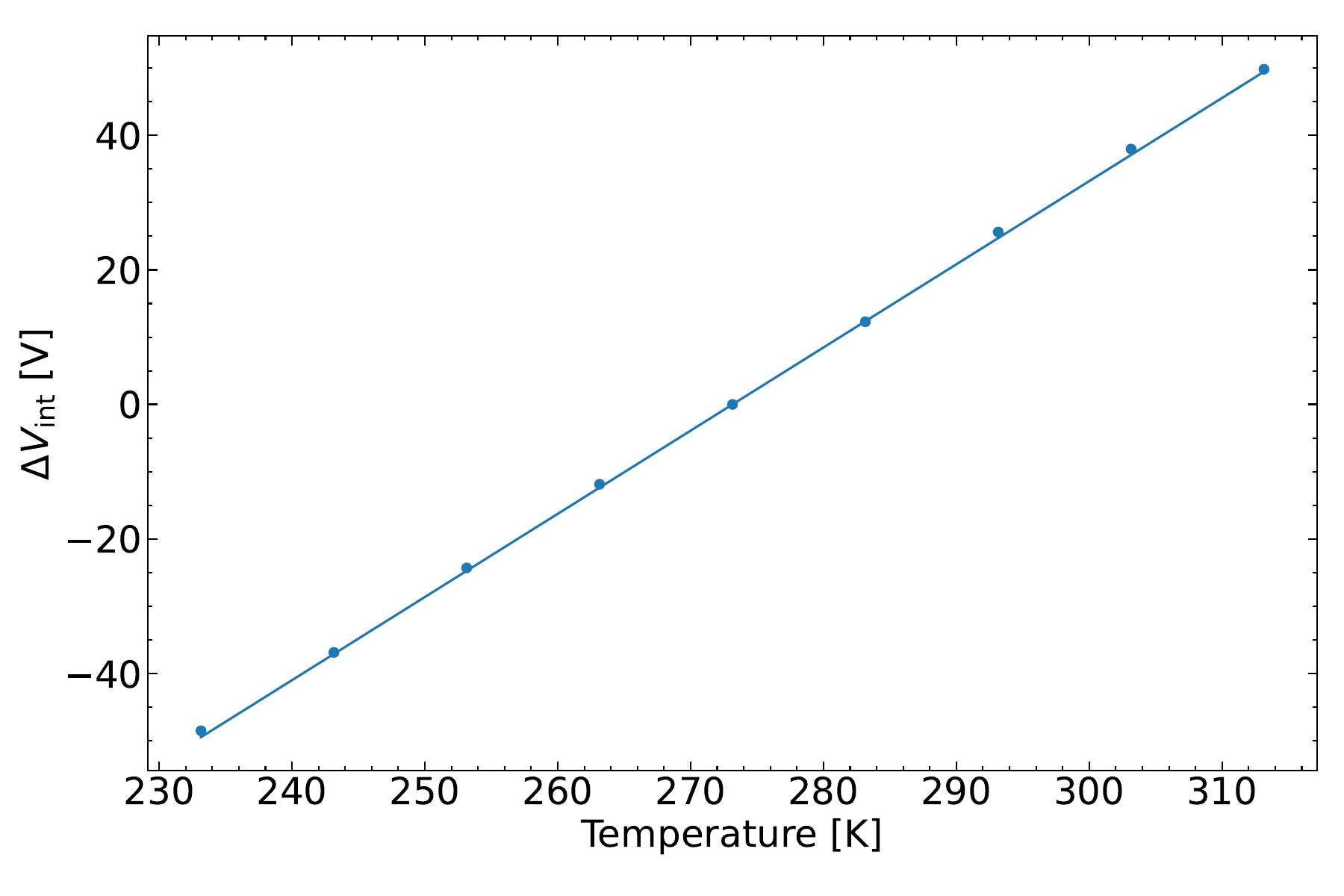}\label{fig:9_is}}
\hfill
\caption{Intrinsic-side bias--temperature equivalence. (a) Measured $\sigma_{\rm int}(V_{\rm b})$ curves at nine temperatures, translated to 273~K for the global fit. (b) Extracted equivalent bias offset $\Delta V_{\rm int}(T)$ with a linear fit.}
\label{fig:int_shift}
\end{figure}

Using the same bias-offset matching procedure as for jitter with $T_{\rm ref}=273$~K, we extract $\Delta V_{\rm int}(T)$ from the nine temperature datasets and obtain the linear fit
\begin{equation}
\Delta V_{\rm int}(T)\approx s_{\rm int}(T-T_{\rm ref}),
\qquad
s_{\rm int}=(1.24\pm0.01)\ {\rm V/K},
\label{eq:shift_int_fit}
\end{equation}
shown in Fig.~\ref{fig:int_shift}(a) and Fig.~\ref{fig:int_shift}(b).

% ----------------------------------------------------------------------
\subsection{Total time resolution}
\label{subsec:tot_shift}

Combining equations~\eqref{eq:sig_quadsum} and \eqref{eq:sigint_G}, the total model can be written as
\begin{equation}
\sigma_t^2(T,V_{\rm b})
=
\sigma_{\rm jitter}^2(T,V_{\rm b})
+\sigma_0^2+\frac{A_G}{G(T,V_{\rm b})}.
\label{eq:total_model}
\end{equation}

As shown above, the temperature dependences of $\sigma_{\rm jitter}$ and $\sigma_{\rm int}$ can both be parameterized by linear equivalent bias offsets, although their slopes are not identical. We therefore consider two reconstruction methods for $\sigma_t$:

\paragraph{Method a: single-offset reconstruction of the total curve}
\begin{equation}
\sigma_{t,\mathrm{recon}}^{(1)}(T,V_{\rm b})
=
\sigma_t\!\left(T_{\rm ref},V_{\rm b}-\Delta V_{\rm tot}(T)\right).
\label{eq:recon_tot}
\end{equation}

\paragraph{Method b: component-wise offset reconstruction}
\begin{equation}
\begin{aligned}
\sigma_{t,\mathrm{recon}}^{(2)}(T,V_{\rm b})
=
\Big[
&\sigma_{\rm jitter}^2\!\left(T_{\rm ref},V_{\rm b}-\Delta V_{\rm jitter}(T)\right)\\
&+\sigma_{\rm int}^2\!\left(T_{\rm ref},V_{\rm b}-\Delta V_{\rm int}(T)\right)
\Big]^{1/2}.
\end{aligned}
\label{eq:recon_split}
\end{equation}

\begin{figure}[!t]
\centering
\includegraphics[width=0.88\linewidth]{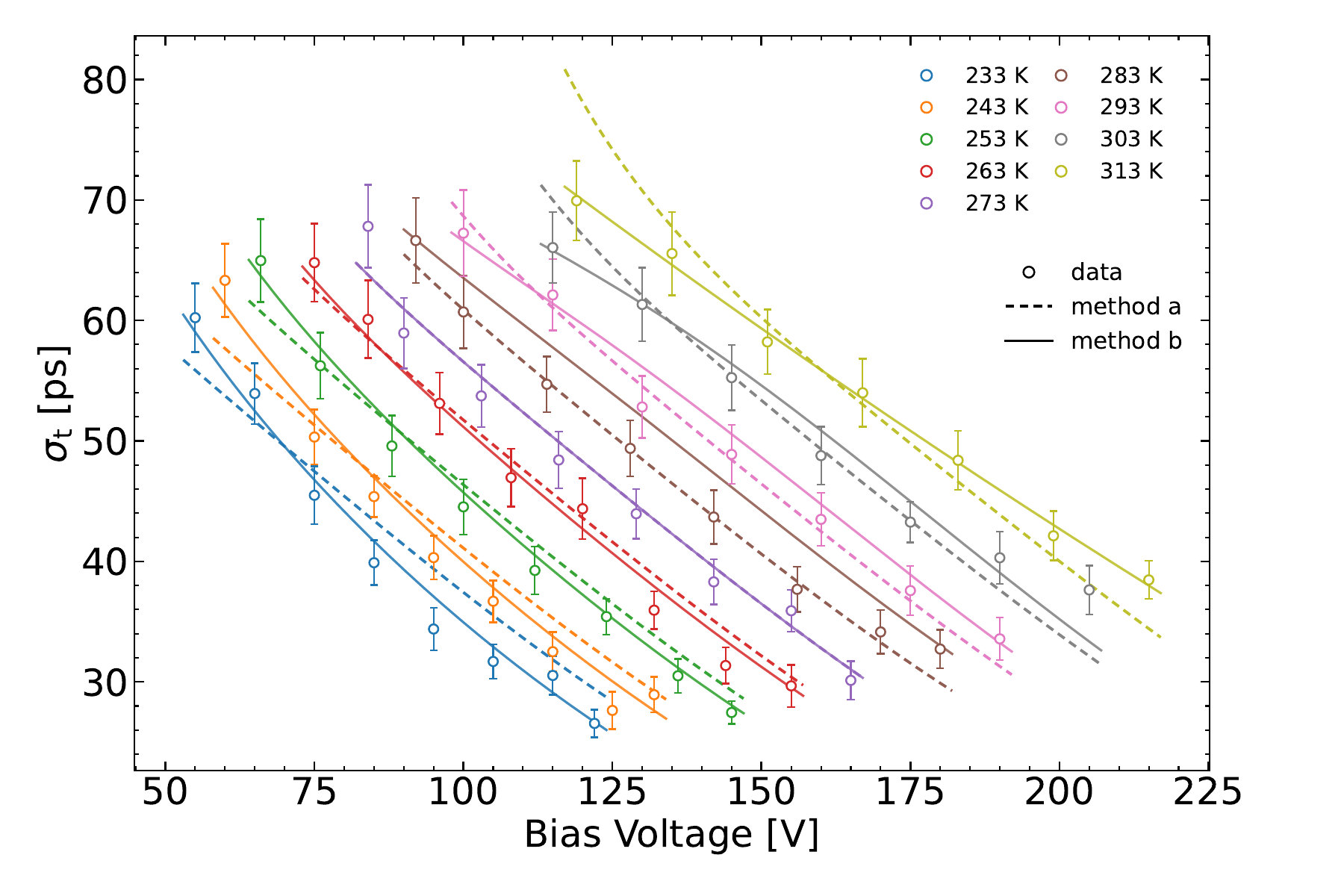}
\vspace{-0.3em}
\caption{Measured total time-resolution curves $\sigma_t (V_b)$ at nine temperatures. The single-offset reconstruction (Method A) is shown as dashed lines, and the component-wise reconstruction (Method B) as solid lines.}
\vspace{-0.2em}
\label{fig:recon_compare}
\end{figure}

\begin{figure}[!t]
\centering
\subfloat[]{\includegraphics[width=0.76\linewidth]{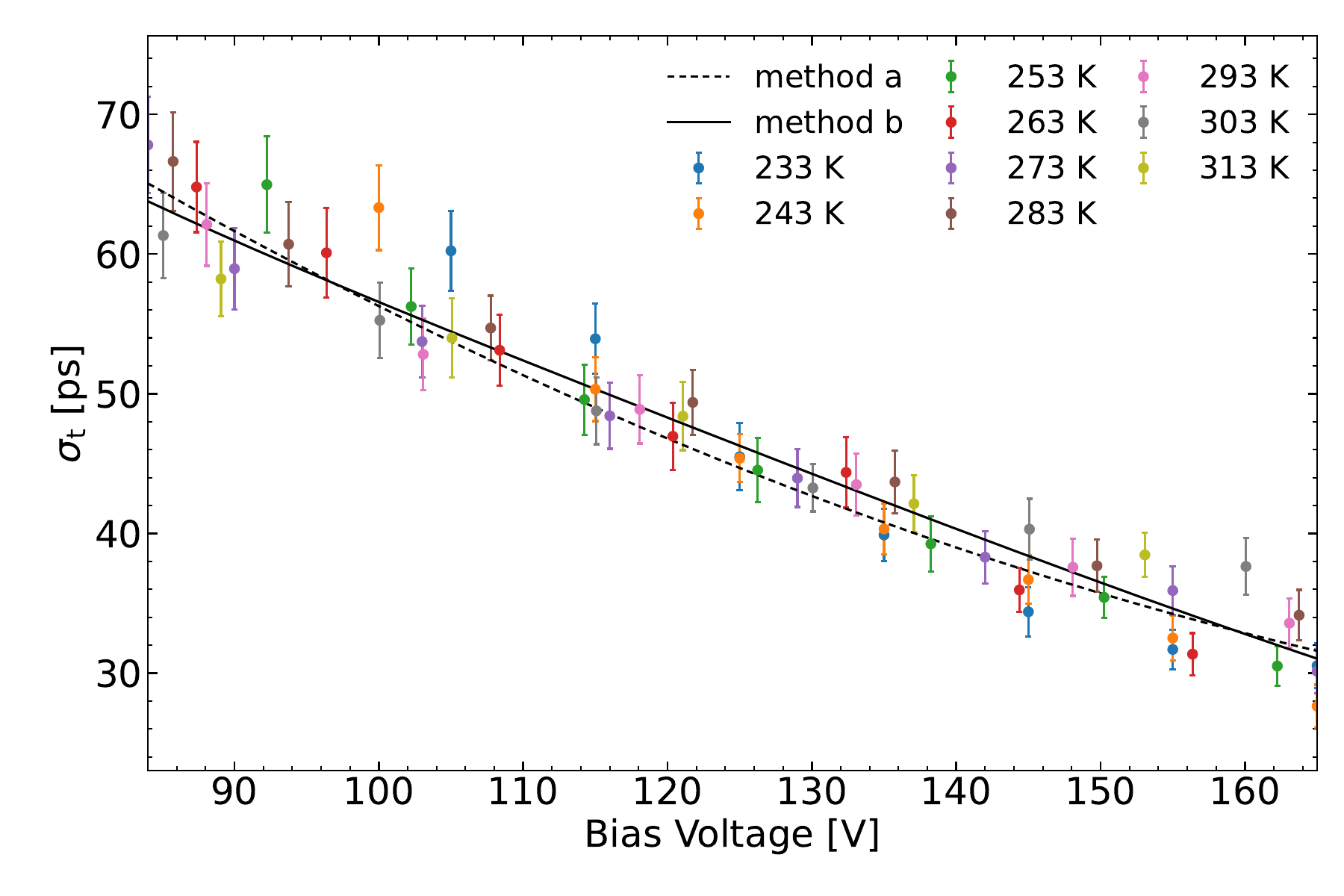}\label{fig:9_tc}}
\vspace{-0.2em}
\subfloat[]{\includegraphics[width=0.76\linewidth]{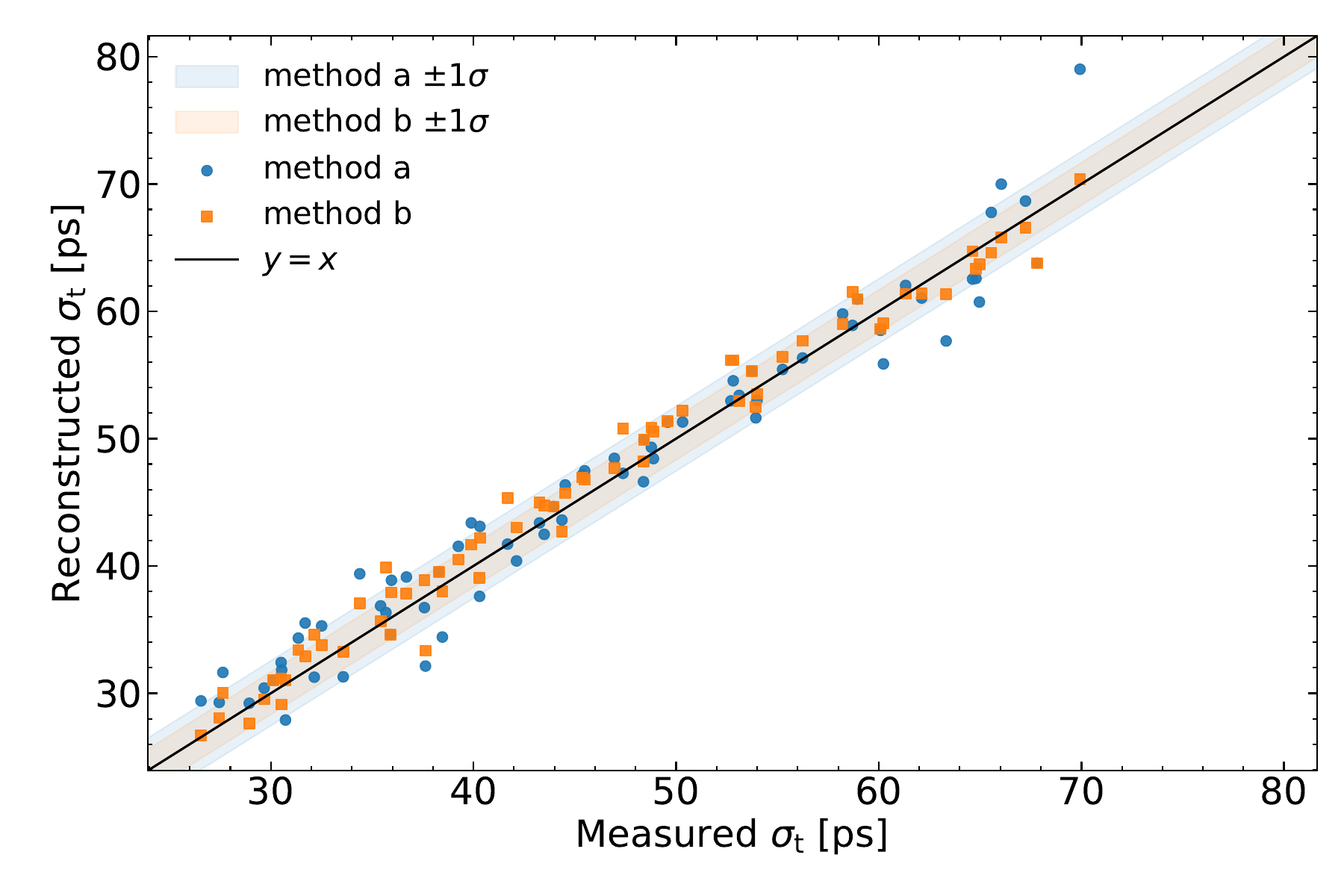}\label{fig:9_ts}}
\vspace{-0.3em}
\caption{Comparison of the two total-time-resolution reconstruction methods. (a) Reconstructed and measured $\sigma_t(V_{\rm b})$ curves at nine temperatures. (b) Scatter of reconstructed versus measured $\sigma_t$.}
\vspace{-0.4em}
\label{fig:tres_9temps}
\end{figure}

The measured total time-resolution curves are shown in Fig.~\ref{fig:recon_compare}, with fittings using both method a (dashed) and method b (solid). Method~a yields an RMSE of $2.56$~ps and a maximum absolute deviation of $9.07$~ps, whereas Method~b reduces these to $1.83$~ps and $4.28$~ps, respectively. Fig.~\ref{fig:tres_9temps}(a) compares the reconstructed and measured $\sigma_t(V_{\rm b})$ curves at all nine temperature points, and Fig.~\ref{fig:tres_9temps}(b) shows the difference between reconstructed and measured time resolution. These results show that, within the studied temperature range, component-wise bias--temperature equivalence provides a more consistent description than a single offset applied to the total curve, corresponding to an improvement of more than 29\% in RMSE.

\FloatBarrier
\section{Conclusion}
This work develops a compact analytical framework for the temperature-dependent gain and timing characteristics of LGADs.

On the gain side, the rectangular gain-layer equivalence (rectGL) reduces the non-uniform multiplication region to a three-parameter device-level model and leads to a first-order bias-compensation relation for constant gain. The framework reproduces the multi-temperature measurements of the IHEP-IME LGADs studied here and transfers well to an independent HPK dataset. In the present datasets, 5--8 bias points at the reference temperature, together with one or two anchor points at each additional temperature, are sufficient to reconstruct the full gain family with good accuracy.

On the timing side, the same bias--temperature-equivalence idea can be applied separately to the jitter and intrinsic components. Their component-wise reconstruction describes the measured time-resolution family more accurately than a single compensation offset applied to the total curve and provides a compact function-level description of multi-temperature timing curves.

Overall, the framework connects bias, temperature, gain, and
timing in an interpretable device-level way and offers a practical route for reduced-density characterization, calibration, and operation of LGAD-based ultrafast timing systems.

\FloatBarrier
\printbibliography
% \bibliographystyle{IEEEtran}
% \bibliography{VTBeta}

\end{document}